\documentclass[prd, twocolumn, nofootinbib,floatfix]{revtex4}

\usepackage{epsfig}

\newcommand{\beq}{\begin{equation}}
\newcommand{\eeq}{\end{equation}}
\newcommand{\beqa}{\begin{eqnarray}}
\newcommand{\eeqa}{\end{eqnarray}}
\newcommand{\om}{\Omega_m}

\newcommand{\wi}{w_\infty}

\def\fun#1#2{\lower3.6pt\vbox{\baselineskip0pt\lineskip.9pt
  \ialign{$\mathsurround=0pt#1\hfil##\hfil$\crcr#2\crcr\sim\crcr}}}

\begin{document} 

\title{Dark Entropy: Holographic Cosmic Acceleration} 
\author{Eric V. Linder} 
\affiliation{Physics Division, Lawrence Berkeley Laboratory,
Berkeley, CA 94720, USA} 
\email{evlinder@lbl.gov}

\begin{abstract} 
In cosmic holography, the fundamental quantity is the degrees of 
freedom on a horizon surface rather than the material contents 
within the volume.  That is, the horizon area and hence cosmological 
expansion rate $H$ is related to the entropy.  Using this as a guide, 
we examine the consequences of an effective dark {\it entropy} 
on the past and future dynamics of the universe.  The results, including 
nonmonotonic expansion behavior, are interesting apart from the theoretical 
motivation. 
\end{abstract} 

\maketitle

\section{Introduction} \label{sec.intro}

Acceleration of the cosmic expansion has profound implications for 
not only the dynamics and contents of the universe, but for 
fundamental physics on both very high energy physics and low 
energy physics scales.  Explanations of the acceleration have 
been sought within new particles and fields, extra dimensions, 
extensions of gravitation theory, and string theory.  More 
frequently, purely ad hoc modifications of the Friedmann 
equations for the expansion dynamics have been proposed. 

To avoid purely phenomenological, unmotivated mutations of 
the Friedmann expansion equation, one might be tempted to 
concentrate on quantities directly related to the physics. 
One example is the geometric dark energy proposed by 
\cite{lingrav} involving the Ricci scalar curvature; by the 
Principle of Equivalence one might expect the spacetime 
curvature to be intimately connected with the acceleration. 

Another possibility is the comoving horizon scale $(aH)^{-1}$, 
a central quantity in early universe inflation.  The time 
change of this quantity explicitly indicates whether the 
universe is accelerating or decelerating, and the horizon 
scale is closely tied to the flatness, structure, and causality 
issues of inflationary cosmology. 

Here we draw our inspiration from holography and consider the 
horizon area $H^{-2}$ as a key quantity.  Recall that 
attempts to resuscitate a positive cosmological constant $\Lambda$ 
as an explanation for the acceleration run into fine tuning and 
coincidence problems \cite{carroll}.  Indeed the scale 
of any vacuum energy density seems generically much larger than 
the $(\sim10^{-3}\,eV)^4$ 
required to explain the observations.  One way out of this is if 
the vacuum energy is not a sum over modes up to some cutoff energy, 
but rather bounded by spacetime geometry.  Such a constraint could follow 
from the holographic principle limiting the number of possible quantum 
states \cite{thomas}.  We do not here assume that but use that holography 
views the horizon surface as 
fundamental and relates it to an entropy, and hence information, bound 
\cite{boussormp}. 

Now, rather than adding 
terms to the Hubble expansion $H^2$ corresponding to new energy 
components in the volume, e.g.\ quintessence, we add terms to its 
reciprocal, the horizon area. 
From holography this denotes adding degrees of freedom, and 
effectively corresponds to new {\it entropy\/} components. 

In \S\ref{sec.dyn} we investigate the dynamical implications of 
this ansatz, including the past and future expansion behaviors. 
\S\ref{sec.cos} investigates tests of the scenario through 
cosmological probes such as distance-redshift and growth factor 
measurements. 
\S\ref{sec.holo} discusses further aspects of the entropy 
interpretation and its place within holographic cosmology. 

\section{Dynamics of Dark Entropy} \label{sec.dyn} 

The basic approach is view the horizon entropy as the central 
quantity and posit a dark sector contribution in addition to the 
usual matter term (we concentrate on the recent and future 
universe).  So 
\beq 
S=S_{\rm matter}+S_{\rm dark}.
\eeq 
Being concerned with the dynamics, we focus on the time dependence 
of these quantities.  We relate this to the expansion 
behavior $H^2$, and observations that depend upon that, through 
the horizon area $A\sim H^{-2}$ and the thermodynamic relation 
$A\sim S$.\footnote{This is not quite standard holography in that 
it is not just $S_{\rm dark}\sim A$.} 

At high redshift (small expansion factor $a$), the universe should 
be matter dominated and we need to recover the matter dominated 
Friedmann equation.  In the future (large $a$), we anticipate that 
the dark sector takes over and perhaps the universe reaches an 
asymptotic deSitter state.  But to retain more generality we allow the 
asymptotic future equation of state to be $\wi$ rather than fixing 
it to the value $-1$. 

Our ansatz therefore becomes 
\beqa
H^2/H_0^2 &=& 1/[\om^{-1}a^3f(a)+\Omega_d^{-1}a^{3(\wi+1)}g(a)] 
\label{eq.h2a} \\ 
&=& \om a^{-3}/\left[f(a)+(\om/\Omega_d)a^{3\wi}g(a)\right], \label{eq.h2}
\eeqa 
where $\om$ is the dimensionless matter density and $\Omega_d$ is an 
effective dark energy density.  
Here the functions $f$ and $g$ represent the transition from 
matter to dark domination, and could also include additional 
factors such as entropy per degree of freedom.  They must satisfy 
the normalization condition at present ($a=1$) of $f(1)+(\om/\Omega_d)g(1)
=\om$. 
At small expansion factor $a\to0$ (high redshift) we take 
$f\to1$ and $g\to0$.  Asymptotically in the future, $a\to\infty$, 
we take $f\to0$ and $g\to1$. 

We have separated out a factor $a^{3\wi}$ from the dark component 
to suggest that asymptotically in the future the expansion rate 
follows that of a component with equation of state ratio $\wi$. 
The comoving horizon shrinks in the future if $\wi<-1/3$. 
For example if $\wi=-1$ then the universe attains a deSitter 
state with constant $H$, constant physical horizon area, and constant 
entropy. 

Defining $\delta H^2$ as the part of the expansion rate not due to 
matter, i.e\ the ``dark'' expansion, we rewrite Eq.\ (\ref{eq.h2}) as 
\beqa 
H^2/H_0^2 &\equiv& \om a^{-3}/J(a) \\ 
\delta H^2 &\equiv& (H^2/H_0^2)-\om a^{-3}=\om a^{-3}\left[J^{-1} 
-1\right]. 
\eeqa 
We can calculate the effective total equation of state of the 
universe and the effective dark equation of state respectively by 
\beqa 
w_{\rm tot} &\equiv& -1-\frac{1}{3}\frac{d\ln H^2/H_0^2}{d\ln a} 
\label{eq.wtoth} \\ 
w &\equiv& -1-\frac{1}{3}\frac{d\ln \delta H^2}{d\ln a}=w_{\rm tot} 
\frac{H^2/H_0^2}{\delta H^2}. \label{eq.wdh} 
\eeqa 

The effective dark equation of state is 
\beq 
w=\frac{1}{3}\frac{af'+a^{3\wi+1}G'+3\wi a^{3\wi}G}{(f+a^{3\wi}G) 
(1-f-a^{3\wi}G)}, 
\eeq 
where $G(a)=G(\infty)\,g(a)=(\om/\Omega_d)g(a)$. 
In the future, $a\to\infty$, we recover $w=\wi$ as long as the 
transition $1-g$ approaches 0 swiftly enough.  Well into the 
matter dominated epoch, $a\to 0$, we find 
\beq 
w(a\to 0)=\frac{1}{3}\frac{af'}{1-f}\to -\frac{1}{3}, 
\eeq 
using a Taylor expansion of $f$, and requiring $g$ to transition 
to 0 sufficiently swiftly in the past.  Under these conditions 
the dark entropy always acts as an accelerating component: 
$w\in [-1/3,\wi]$. 

The effective total equation of state is 
\beq 
w_{\rm tot}=\wi+\frac{1}{3}\frac{af'-3\wi f+a^{3\wi+1}G'}{f+a^{3\wi} 
G}. 
\eeq 
As $a\to 0$, i.e.\ well into the matter dominated epoch, we find 
$w_{\rm tot}=0$, as expected.  As $a\to\infty$, $w_{\rm tot}\to\wi$. 

We discuss further interpretation of quantities $f$ and $g$ in 
\S\ref{sec.holo}. 

One possible realization of the transition functions that satisfies 
all the above conditions is 
\beqa 
f &=& e^{-a/a_m} \\ 
G &=& (\om-e^{-1/a_m})e^{a_d(1-1/a)}=(\om/\Omega_d) e^{-a_d/a}. 
\eeqa 
Note that $g=e^{-a_d/a}$ and $G\sim e^{-a_d z}$.  
Here $a_m$ and $a_d$ characterize the times of the transitions; we 
keep them separate for generality.  Note that we implicitly assumed 
that $\wi<0$ (indeed we might expect that $\wi=-1$ so as to attain a 
deSitter state; see \cite{boussormp} for a discussion of holography 
and deSitter spacetimes), but one could allow for $\wi>0$ so as to treat 
models of a collapsing universe in the future where $\wi\to +\infty$ 
(equivalent to a scalar field where the potential energy is negative 
enough to balance the kinetic energy: $V+K=0$, cf.\ the linear 
potential doomsday model \cite{linde} or cyclic model 
\cite{turokstein}). 

We find that the above ansatz gives a rich phenomenology for the 
expansion dynamics, much richer than for quintessence, say.  Even 
fixing $\om=0.28$ and $\wi=-1$ allows a 
variety of behavior as one varies $a_m$ and $a_d$.  This can include 
a nonmonotonic Hubble parameter $H(a)$, negative $\delta H^2$, and 
generally nonmonotonic $w(a)$ and $w_{\rm tot}(a)$.  None of these 
are necessarily physically pathological, and it will be up to observations 
involving expansion history to constrain the parameter values.  
\S\ref{sec.cos} addresses this issue. 

\begin{figure}[!t]
\begin{center}
\psfig{file=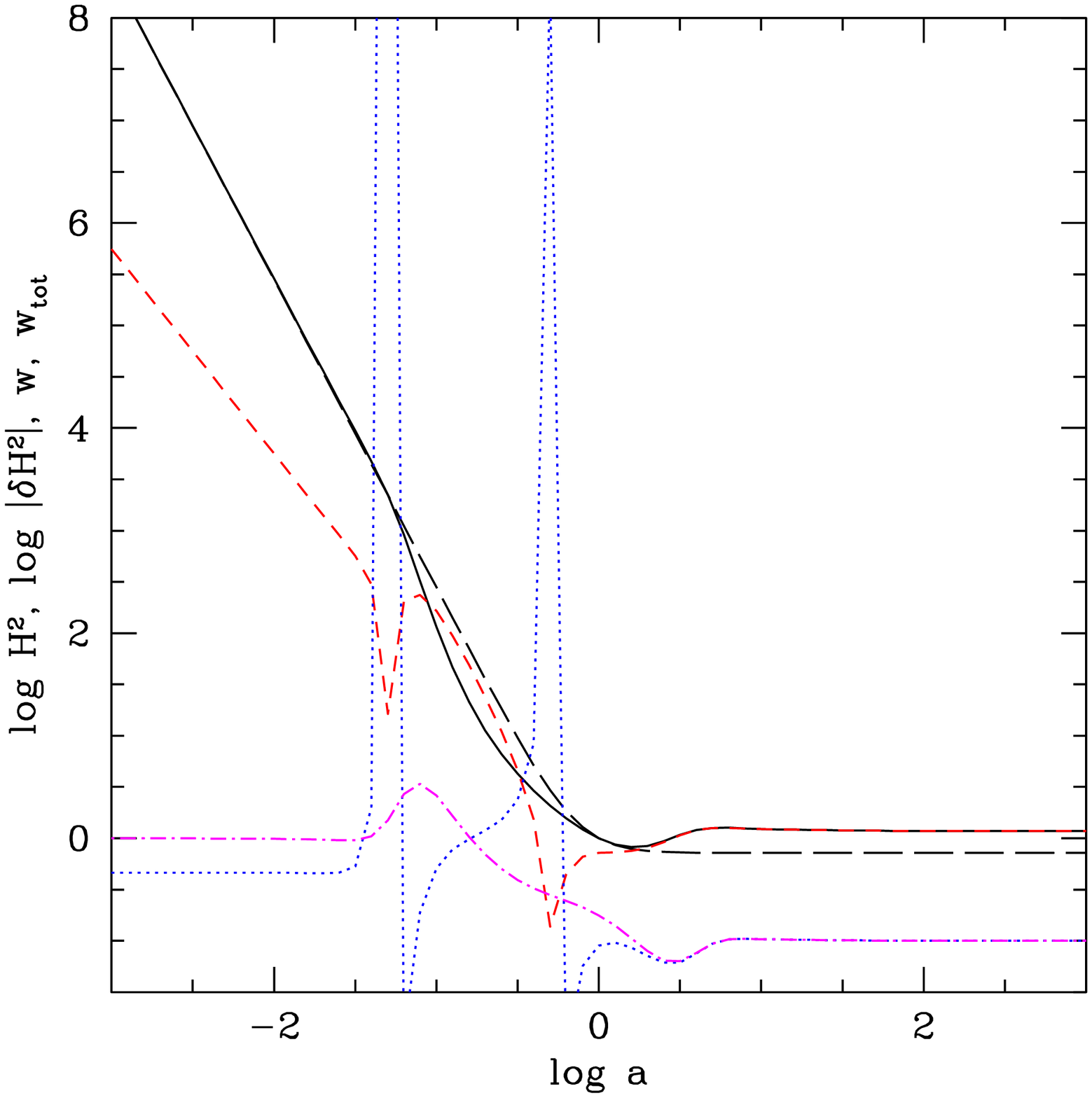,width=3.4in}
\psfig{file=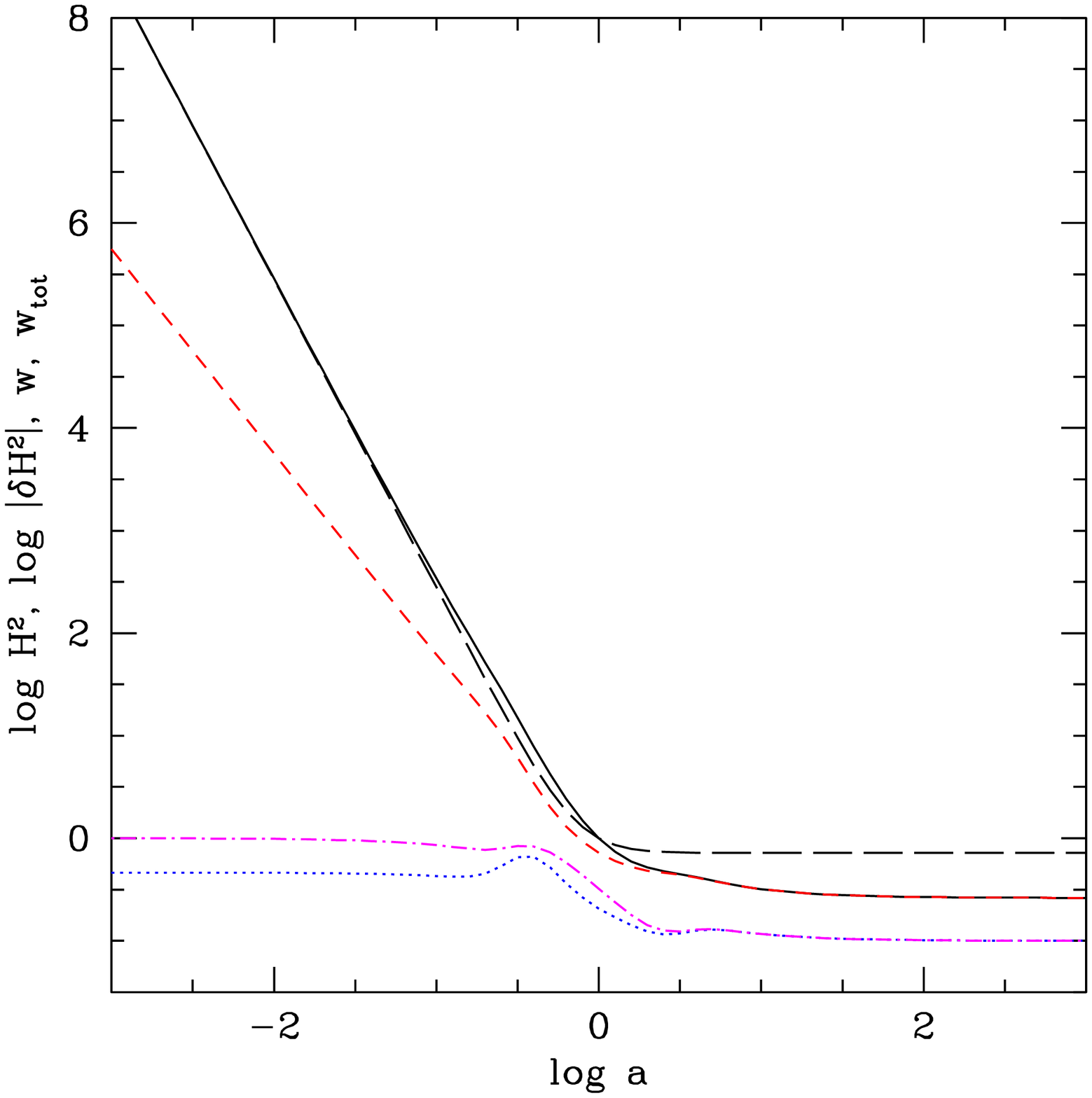,width=3.4in} 
\caption{Dark entropy models exhibit a variety of expansion dynamics
behaviors.  The expansion rate squared $H^2$ (black, solid curve) 
can be nonmonotonic, in contrast to the cosmological constant (black 
long dashed) or quintessence case.  The deviation from matter 
dominated expansion, $\delta H^2$ (red, dashed), can be substantial, 
or give a negative contribution (between cusps).  The total equation 
of state (magenta, dot dashed) and effective dark equation of state 
(blue, dotted) similarly have nonmonotonic behaviors. 
Top plot has $(a_m,a_d)=(0.5,0.5)$; bottom has $(0.5,2)$.  
}
\label{fig.models12}
\end{center}
\end{figure}

\begin{figure}[!t]
\begin{center}
\psfig{file=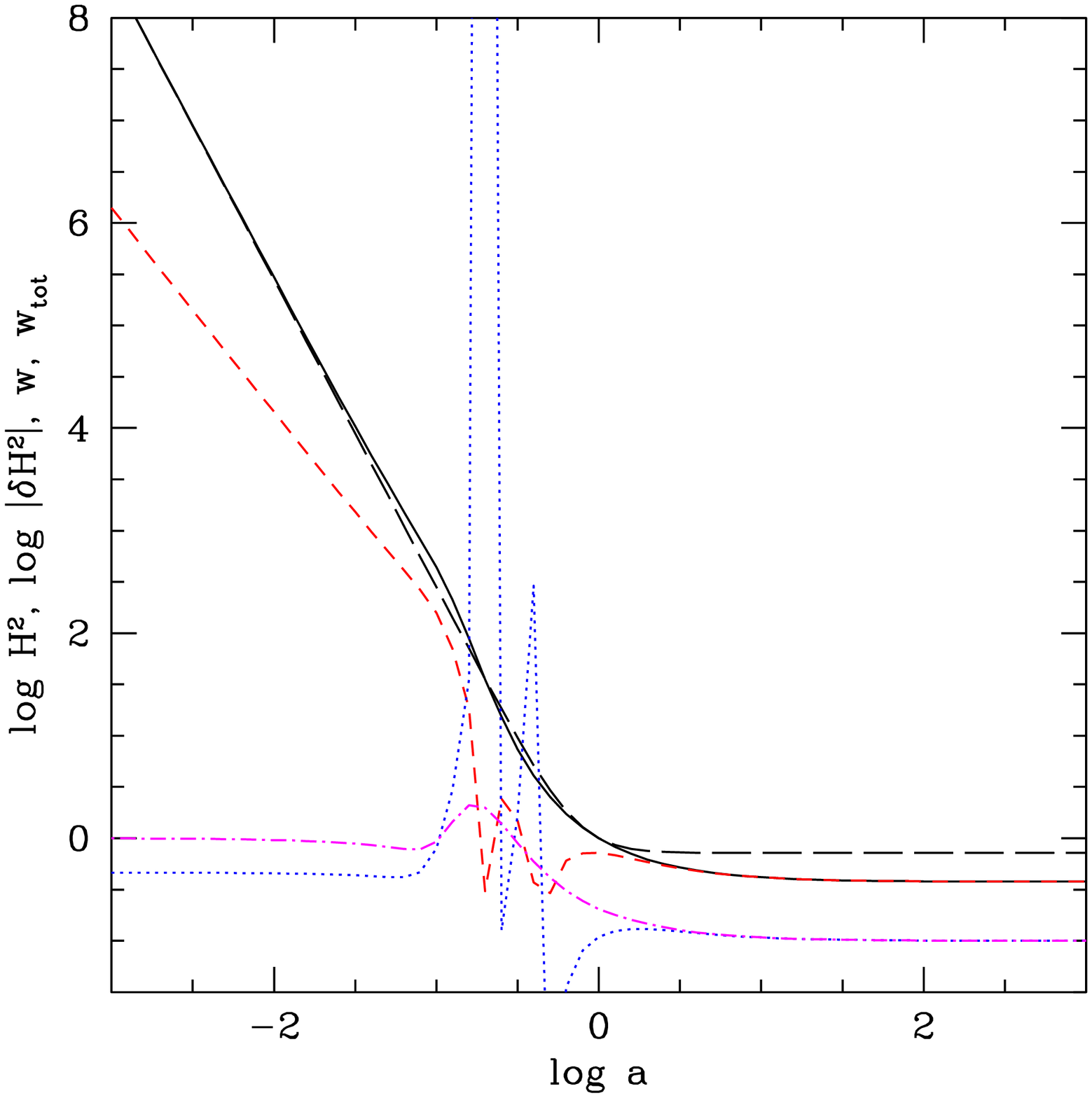,width=3.4in}
\psfig{file=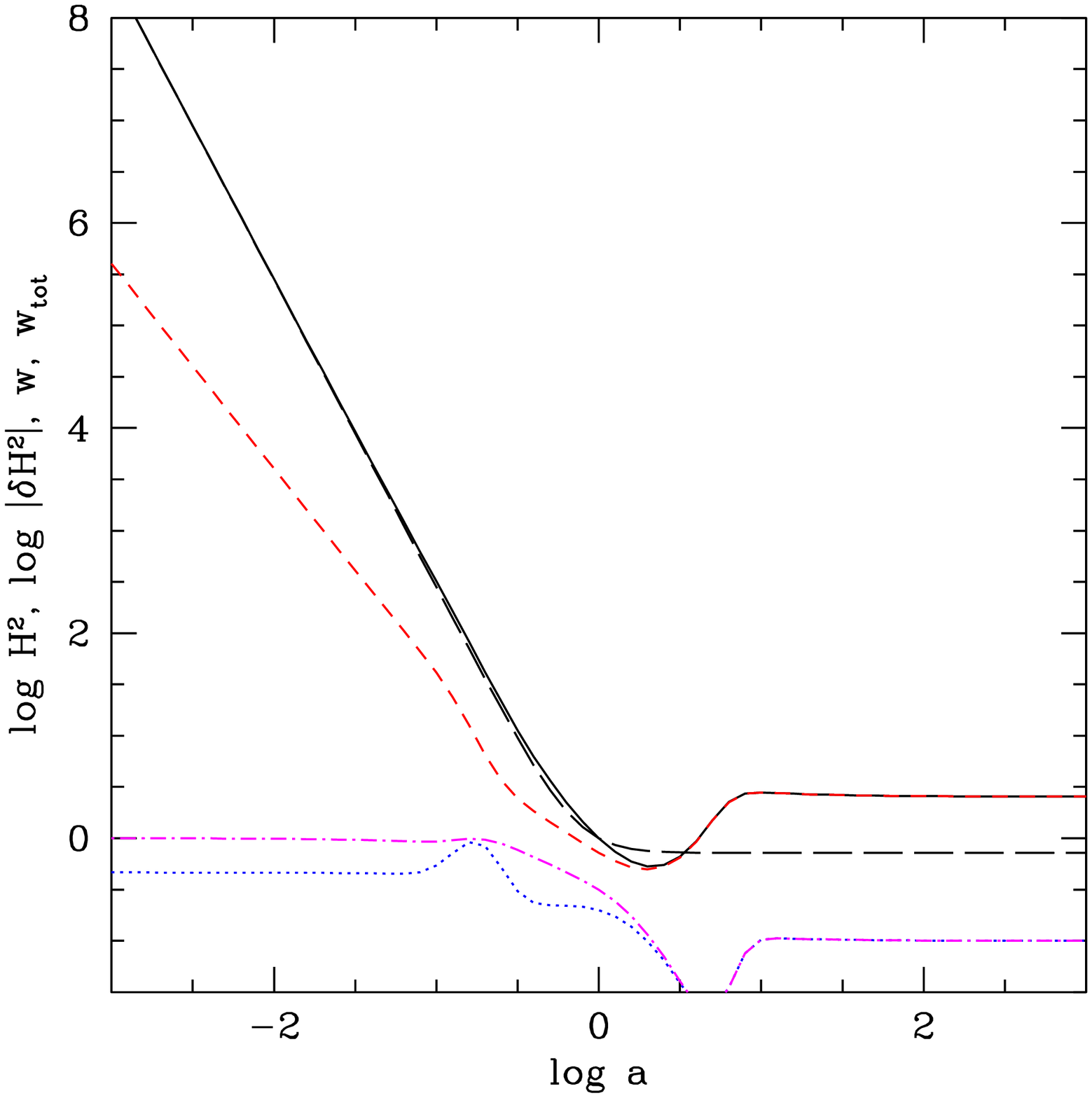,width=3.4in}
\caption{As Fig.\ \ref{fig.models12}, but here the 
top plot has $(a_m,a_d)=(0.2,1)$; bottom has $(0.7,1)$.
}
\label{fig.models34}
\end{center}
\end{figure}

Figures \ref{fig.models12} and \ref{fig.models34} illustrate the 
variety of expansion and equation of state behaviors.  At high 
redshift (small $a$), the expansion rate squared $H^2$ (solid, black 
curve) follows the matter dominated form; the deviation $\delta H^2$ 
(dashed, red curve) from this is several orders of magnitude smaller. 
The total equation of state of the universe (dot dashed, magenta curve) 
is $w_{\rm tot}=0$ then, as for matter domination, and the effective 
dark equation of state (dotted, blue curve) from the dark entropy 
goes to $w=-1/3$.  As the dark entropy gains importance, the expansion 
rate differs from that of a flat universe with a cosmological 
constant (long dashed black curve) and the equations of state vary 
as well.  

The behaviors not only deviate from those in a cosmological constant 
universe but from quintessence cosmologies generally, and can be 
nonmonotonic in any of the quantities.  Furthermore, $\delta H^2$ can 
go negative; though the figures plot the absolute value, one can 
recognize such a region as that lying between the cusps, e.g.\ at 
$\log a=-1.3$ to $-0.3$ in the top plot of Fig.\ \ref{fig.models12}. 
When $\delta H^2$ crosses through zero, $w$ goes to infinity, as is 
clear from Eq. (\ref{eq.wdh}).  As $\delta H^2$ reaches its most 
negative value, $H^2$ gets smaller and steeper, causing $w_{\rm tot}$ to 
rise (as per Eq.\ \ref{eq.wtoth}), giving a hump and possibly 
$w_{\rm tot}>0$.  Conversely, both $w$ and $w_{\rm tot}$ can fall 
below $-1$ if the balance between matter and dark entropy causes 
$\delta H^2$ to experience an epoch where it increases with time. 
In this case the overall expansion rate $H^2$ can also increase 
with time, giving a nonmonotonic dip and rise behavior. 

Asymptotically in the future the expansion rate settles down to 
a constant, deSitter state, and both the effective dark and total 
equations of state become $-1$ (assuming $\wi=-1$, as we do for 
the rest of the paper). 

Holding $\om=0.28$, we see that as the dark dominance comes earlier 
(decreasing the parameter $a_d$), one
tends to move the positive bump in $w_{\rm tot}$ earlier, obtain a
period of negative $\delta H^2$, and create strong oscillations in $w$. 
It also tends to make $w_{\rm tot}$ more negative today. 
Decreasing the matter dominance (smaller $a_m$) makes the bump in
$w_{\rm tot}$ more positive. 
We could systematically study the phase space and seek out those 
regions with particular properties, e.g.\ smooth behavior in $w$ 
or only a small bump to $w_{\rm tot}$.  In the next section we 
address this from an observational basis. 

\section{Cosmological Probes of Dark Entropy} \label{sec.cos} 

\begin{figure}[!hbt]
\begin{center}
\psfig{file=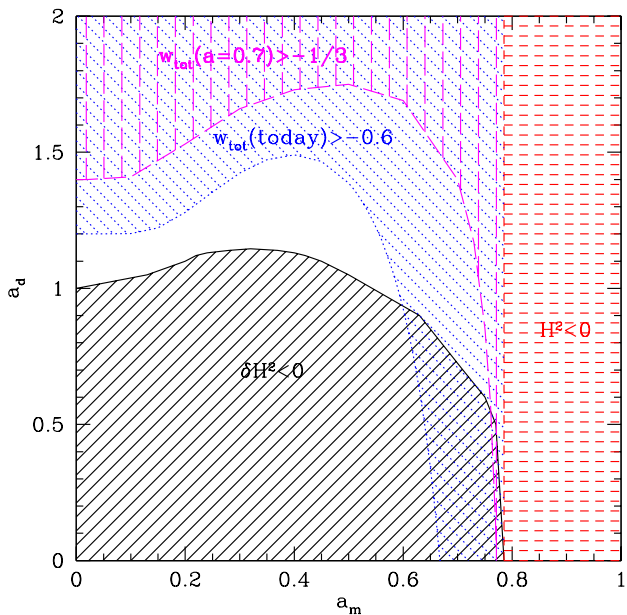,width=3.4in}
\caption{Limits on the transition parameters $a_m$ and $a_d$ for 
various physical conditions, leaving only a narrow window of phase space. 
}
\label{fig.amad}
\end{center}
\end{figure}

The parameters $a_m$, $a_d$ describing the transition behaviors of 
the matter and dark components determine the character of the expansion 
rate and equations of state.  By placing various constraints upon them 
we can restrict the allowed phase space. 
To assure expansion -- positive $H(a)$ -- we require $f+Ga^{3\wi}>0$. 
The condition for this to hold at all redshifts is $G>0$ or 
$a_m<1/\ln\om^{-1}$.  For 
$\om=0.28$ this implies $a_m<0.785$.  One might require that the 
expansion be accelerating, i.e.\ the total equation of state 
$w_{\rm tot}<-1/3$, below some redshift $z_{acc}$, or that the present 
dark equation of state be less than some value.  For example, if $a_m=0.5$ 
then to have acceleration today such that $w_{\rm tot}<-0.6$ (for 
quintessence this would correspond to $w<-0.83$) would require $a_d<1.39$. 
These constraints are shown in Fig.\ \ref{fig.amad}, leaving open only 
a narrow window of phase space.  Note that the allowed region almost entirely 
possesses $a_d>1$, meaning that true dark entropy dominance is restricted 
to not occur in the past in cosmologies that look like ours. 

\begin{figure}[!t]
\begin{center}
\psfig{file=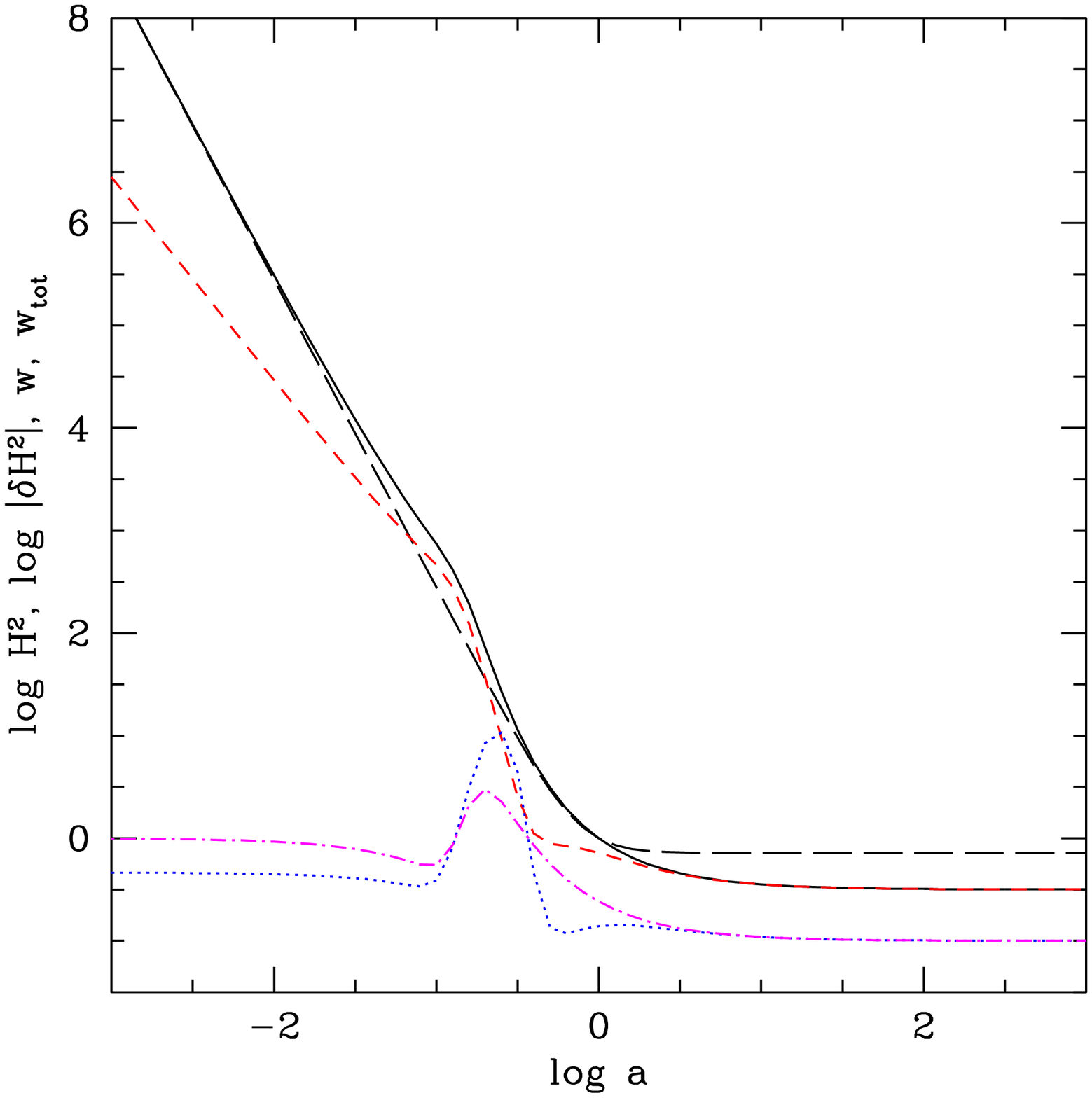,width=3.4in}
\psfig{file=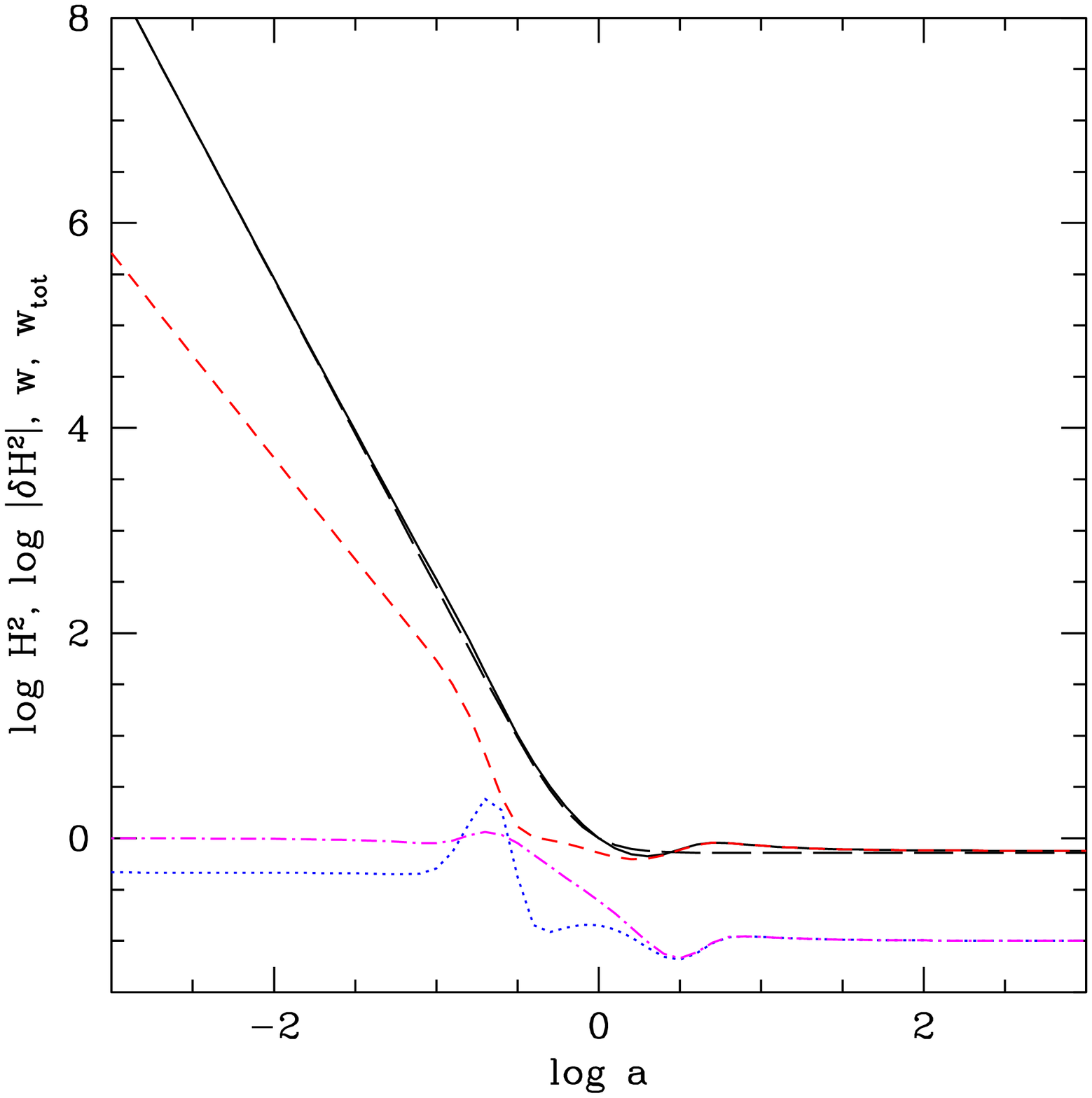,width=3.4in} 
\caption{Expansion histories and equation of state histories, 
as in Fig.\ \ref{fig.models12}, of 
selected models in the allowed phase space.  Top plot has 
$(a_m,a_d)=(0.1,1.15)$; bottom has $(0.55,1.15)$. 
}
\label{fig.pars12}
\end{center}
\end{figure}

\begin{figure}[!ht]
\begin{center}
\psfig{file=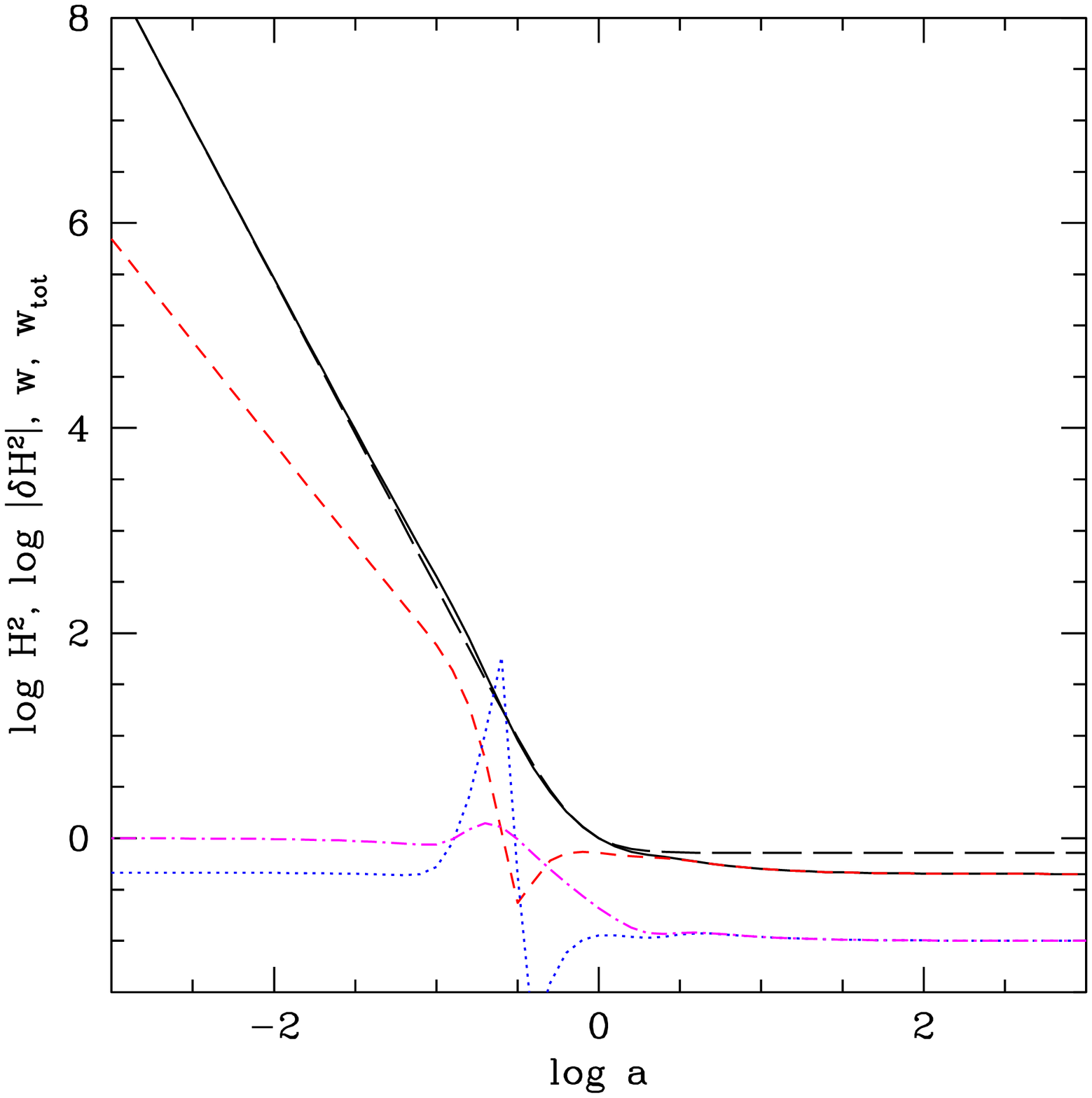,width=3.4in}
\psfig{file=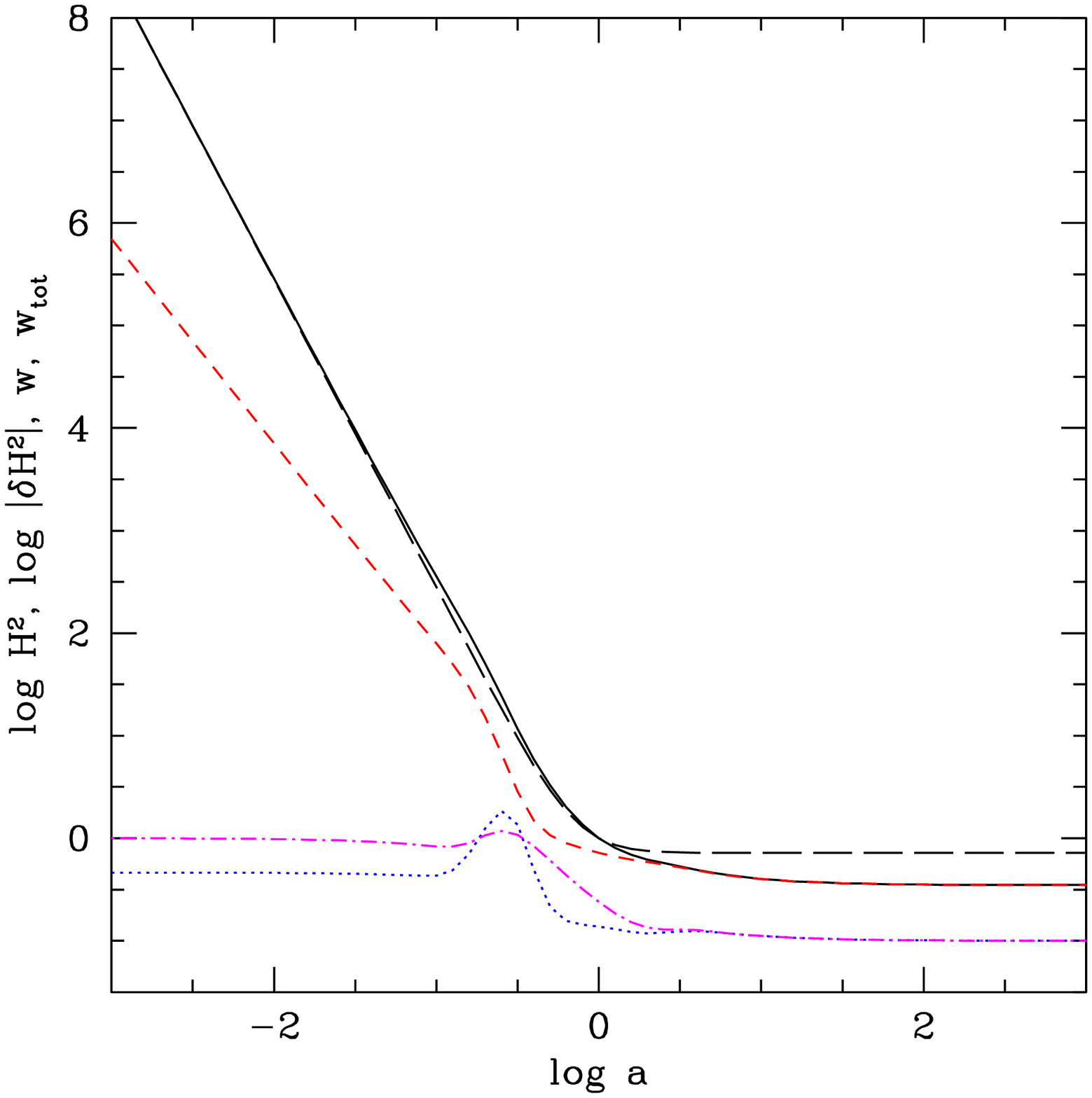,width=3.4in}
\caption{Expansion histories and equation of state histories, 
as in Fig.\ \ref{fig.models12}, of
selected models in the allowed phase space.  Top plot has
$(a_m,a_d)=(0.4,1.15)$; bottom has $(0.4,1.40)$.
}
\label{fig.pars34}
\end{center}
\end{figure}

We select points in the $a_m-a_d$ phase space allowed by all the 
conditions and plot their expansion and equation of state histories 
in Figs.\ \ref{fig.pars12}-\ref{fig.pars34}.  Note that it is not 
obvious that $\delta H^2<0$ is a region to be shunned necessarily 
in and of itself; however, its effect on the equations of state is 
sufficient to cause difficulties upon comparison to observations 
of distances and growth of matter perturbations.  Restricting to 
$\delta H^2>0$ also implies that for models in the allowed region one has 
$w_{\rm tot}({\rm today})>-0.68$ or for a quintessence model $w_0>-0.95$. 

Several of these figures possess expansion histories that look 
(on the log-log plots) extremely 
similar to the flat, matter plus cosmological constant case (long dashed 
curves).  However the equations of state evolutions have definite 
nonmonotonic features.  So it is interesting to consider how these 
dark entropy models would compare to cosmological constant or dark 
energy universes under current or future cosmological observations. 

\begin{figure}[!h]
\begin{center}
\psfig{file=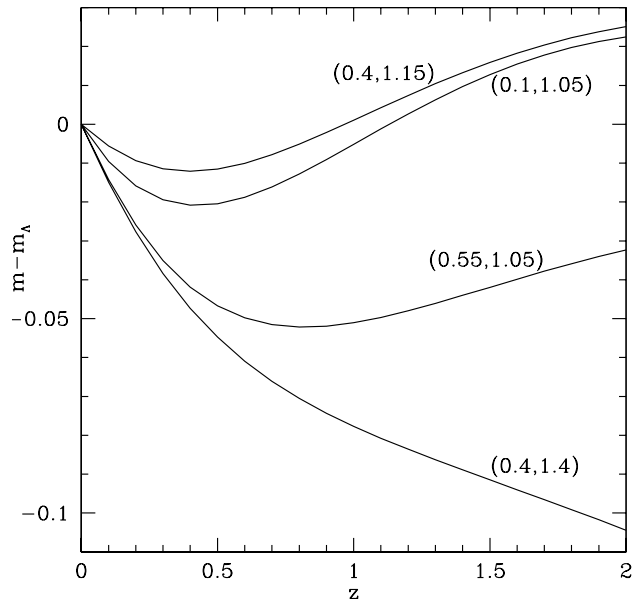,width=3.4in}
\caption{Hubble diagram, or magnitude-redshift relation, of 
selected models in the allowed phase space relative to the relation 
for a cosmological constant universe with the same matter density. 
}
\label{fig.entrom}
\end{center}
\end{figure}

Figure \ref{fig.entrom} investigates this in terms of the 
Hubble diagram, or magnitude-redshift relation, comparing the 
quantity $m=5\log d_l=5\log[(1+z)\int dz/H]$ in the dark entropy 
model parametrized by $(a_m,a_d)$ to the cosmological constant case. 
One sees that models in the allowed parameter space of Fig.\ \ref{fig.amad} 
are within 0.1 mag (5\% in distance) of the cosmological constant behavior, 
about the current discrimination level.  Some models lie within 0.02 mag. 
(Since one is free to adjust the unknown absolute magnitude scale, known 
as the ${\cal M}$ parameter, in fact the deviation of the models is 
one half the difference between the most positive and most 
negative deviations.) 

Some of the activity in the dark entropy expansion history happens at 
redshifts $z>2$ ($\log a<-0.5$), for example the bump to $w_{\rm tot}>0$. 
One might expect this to have an effect on growth of matter perturbations 
and the evolution of large scale structure.  A summary of 
linear perturbation theory applied to universes with nonstandard expansion 
rates and equations of state appeared in \cite{linjen}.  As pointed out 
there, contrary to oft repeated lore, an equation of state $w_{\rm tot}>0$ 
does not hurt structure growth.  Growth is suppressed for the radiation 
dominated epoch ($w_{\rm tot}=1/3$) not because of the equation of state -- 
note the expansion rate and hence Hubble drag is slower the higher 
$w_{\rm tot}$ -- but because of the reduced source term from the matter. 
So growth will be slower than the matter dominated case at a given epoch 
if matter is not completely dominant, and if the total equation of state 
is negative. 

Explicitly, we can write the linear perturbation theory growth equation 
as 
\beqa 
G''&+&a^{-1}G'\left(\frac{7}{2}-\frac{3}{2}w_{\rm tot}\right) \nonumber \\ 
&+&\frac{3}{2}a^{-2}G 
\left(1-w_{\rm tot}-\frac{\om a^{-3}}{H^2/H_0^2}\right)=0, 
\label{eq.grow}
\eeqa 
where $G=\delta/a$, prime denotes $d/da$, and 
$\delta\equiv\delta\rho_m/\rho$ is the 
fractional matter density 
perturbation.  We can solve this for dark entropy models using Eqs.\ 
(\ref{eq.h2}) and (\ref{eq.wtoth}). 

\begin{figure}[!h]
\begin{center}
\psfig{file=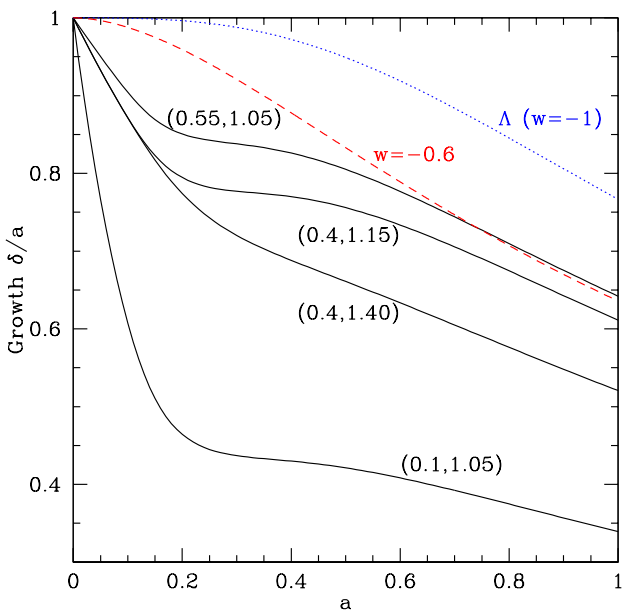,width=3.4in}
\caption{Growth history in linear perturbation theory of 
the same models in the allowed phase space as for Fig.\ \ref{fig.entrom}. 
Also plotted is the growth behavior for cosmological constant and 
quintessence ($w=-0.6$) universes with the same matter density.  The 
growth behavior has several interesting features predicted by the 
variation of the expansion rate or equivalently the nonmonotonic total 
equation of state. 
}
\label{fig.groent}
\end{center}
\end{figure}

The growth behavior is illustrated in Fig.\ \ref{fig.groent} for the 
same models as in the Hubble diagram of Fig.\ \ref{fig.entrom}.  A pure 
matter (Einstein-deSitter) universe would have $\delta/a=1$ for all $a$. 
We clearly see the early suppression of growth, much stronger than for 
the cosmological constant case, due to the reduction in the matter source 
term, i.e.\ the matter is not the only contribution to the effective 
total density $\rho$.  While the dark contribution $\delta H^2$ is an 
order of magnitude smaller than the matter contribution at $a=0.01$, 
this is still much more than the cosmological constant contribution 
would have been: five and a half orders of magnitude smaller than the 
matter density.  So the matter growth lags below that of either the 
pure matter or matter plus cosmological constant case. 

At the time of the $w_{\rm tot}>0$ bump, the growth virtually stops 
declining, as the Hubble drag is reduced (see the $G'$ term in Eq.\ 
\ref{eq.grow}).  Note that writing the growth equation explicitly in 
terms of $w_{\rm tot}$ makes this physics much clearer: increasing 
$w_{\rm tot}$ aids growth and conversely growth is suppressed when 
the total equation of state becomes negative (a more accelerating 
universe).  Closer to the present the matter becomes even less 
dominant and the equation of state goes negative so growth is shut 
down. 

The total growth in the most favorable of the models considered is 
about 16\% less than in the cosmological constant case, and comparable 
to a quintessence universe with $w=-0.6$.  
However, there is still 
some dispute about the amplitude of the mass power spectrum today, described 
by the parameter $\sigma_8$.  If large scale structure data interpreted 
in a cosmological constant universe led to $\sigma_8=0.9$, then the 
top dark entropy curve would correspond to $\sigma_8=0.76$.  This is 
low, but some measurements even lie around 0.7.  For a rigorous comparison 
one should consider the cosmic microwave background power spectrum 
taking into account the difference of the expansion history since the 
last scattering surface, and the matter power spectrum including 
the nonlinear effects entering into $\sigma_8$ determination.  

Furthermore, note that the slopes of the curves are very similar 
near the present.  Since many observations of the growth of structure 
actually involve the gravitational potential $\Phi(a)=G(a)/G(1)$, 
then these would see little difference in the models, out to $z=1$ 
at least where the slopes change.  And slightly different choices 
of $a_m$ and $a_d$ might allow us to improve the agreement of the 
growth history.

\section{Holography and Entropy} \label{sec.holo}

This section speculates on various aspects of the interpretation 
of dark entropy and the relation to cosmic holography.  In classical 
thermodynamics the entropy has various stability conditions related 
to its derivatives, e.g.\ $\partial S/\partial V$ where $V$ is the 
volume.  We should not be surprised if the ansatz here does not follow 
them since we know that gravitating systems are not stable (masses 
aggregate, gravitational systems have a negative specific heat), 
but the form $\partial S/\partial V$ does play 
a role.  Since the volume grows with scale factor as $a^3$, we are 
really concerned with the evolution of $S(a)$.  Fischler \& Susskind 
\cite{fischlersusskind} have discussed this, showing the relationship 
to an effective equation of state within the holographic view. 

It is interesting to examine further the role of the transition 
parameters $f$ and $g$.  They were introduced in order to obtain 
simultaneously the correct asymptotic limits of both the entropy 
and the expansion behavior.  But the following reasoning is suggestive 
of a more basic role.  Write the Hubble parameter for a dark energy 
(here cosmological constant) universe as 
\beqa 
H^2/H_0^2&=&\left[\frac{1}{\om a^{-3}+\Omega_d}\right]^{-1} \\ 
&=&2\left[\om^{-1}a^3  
\frac{\om a^{-3}}{\om a^{-3}+\Omega_d}+\Omega_d^{-1}\frac{\Omega_d}{\om 
a^{-3}+\Omega_d}\right]^{-1}. 
\eeqa 
Now using $H^2\sim S^{-1}$ we see the bracketed quantity bears a 
strong resemblance to our ansatz from Eq. (\ref{eq.h2a}): 
\beq 
S=\om^{-1} a^3 f(a)+\Omega_d^{-1} g(a), 
\eeq 
where now $f$ and $g$ count the proportion of energy in each component, 
e.g.\ serve as fractional degrees of freedom.  Of course we obtain 
$f(a\to0)=1$, $f(a\to\infty)=0$, $g(a\to0)=0$, $g(a\to\infty)=1$ as 
desired.  While suggestive in relating the transition parameters $f$ 
and $g$ to degree of freedom counters, this is only a rough analogy 
since we emphasize that we do not assume any physical energy component 
``dark energy''; rather we have only dark entropy. 

One can define an effective cosmological constant since in 
the asymptotic deSitter state at $a=\infty$, the entropy is 
$S_\infty=3\pi/\Lambda$ \cite{boussormp}.  We have 
\beq 
H_\infty^2/H_0^2=\Omega_d=[\om/(\om-e^{-1/a_m})]\,e^{-a_d}, 
\eeq 
so $\Lambda\sim H_\infty^2$ is of order one, times $H_0^2$, 
or smaller without fine tuning $a_d$ 
(but this follows in part from our requiring a universe like we observe).

\section{Conclusion} 

In searching for new physics to explain the acceleration of the 
expansion of the universe, one can adopt 1) new, and often not well 
defined, components of the energy density, 2) new forms of the gravitation 
equations, or 3) new geometric structures of spacetime.  Here we postulated 
the last route, exploring some implications of the horizon area and its 
associated entropy. 

The picture appears consistent and a real seed of motivation exists 
from the principle of holography. 
While the specific ansatz may not be significantly less ad hoc than 
quintessence models or modifications of the Friedmann equations of expansion 
it does lead to a rich variety of expansion histories of the universe. 
The implications include features of expansion rates and equations of state 
nonmonotonic with scale factor -- an important broadening of physical 
possibilities. 
Furthermore we have shown that the parameter space can be constrained 
by cosmological observations. 

But in addition to fundamental physics connections, one of the 
most significant virtues of the dark entropy scenario is that it is 
distinct from dark energy, including the cosmological constant $\Lambda$. 
That is, there is no limit of the parameters, no corner of phase space, 
that approaches the cosmological constant.  This is in stark contrast 
to almost all viable dark energy potentials or modifications of gravity. 
This property makes dark entropy, aside from other motivations, 
an important benchmark model for 
comparing to observations and understanding the fundamental physics. 

\acknowledgments 

I acknowledge the Aspen Center for Physics for hospitality and a 
stimulating atmosphere.  I also thank Raphael Bousso and Scott Thomas for 
useful discussions. This work has been supported 
in part by the Director, Office of Science, US Department of Energy under 
grant DE-AC03-76SF00098.


\begin{thebibliography}{99}

\bibitem{lingrav}
  E.V.\ Linder, ``Probing Gravitation, Dark Energy, and Acceleration'', 
  Phys.\ Rev.\ D 70, 023511 (2004) [astro-ph/0402503] 

\bibitem{carroll} 
  S.M.\ Carroll, ``The Cosmological Constant'', 
  Living Rev.\ Rel.\ 4, 1 (2001) [astro-ph/0004075] 

\bibitem{thomas} 
  S.\ Thomas, ``Holographic Vacuum Energy'', [hep-th/0010145]

\bibitem{boussormp} 
  R.\ Bousso, ``The Holographic Principle'', Rev.\ Mod.\ Phys.\ 74, 
  825 (2002) [hep-th/0203101] 

\bibitem{linde} 
  R.\ Kallosh et al., ``Observational Bounds on Cosmic Doomsday'', 
  JCAP 0310, 015 (2003) [astro-ph/0307185] 

\bibitem{turokstein} 
  P.J.\ Steinhardt and N.\ Turok, ``The Cyclic Model Simplified'', 
  [astro-ph/0404480]

\bibitem{linjen}
  E.V.\ Linder and A.\ Jenkins, ``Cosmic Structure Growth and Dark 
  Energy'', MNRAS 347, 909 (2004) [astro-ph/0304509] 

\bibitem{fischlersusskind} 
  W.\ Fischler and L.\ Susskind, ``Holography and Cosmology'', [hep-th/9806039]

\end{thebibliography}
\end{document}